# Increased Cycling Efficiency and Rate Capability of Copper-coated Silicon Anodes in Lithium-ion Batteries

Vijay A. Sethuraman,[a,b,*,1] Kristin Kowolik,[a,2] and Venkat Srinivasan[a,*]

[a]Environmental Energy Technologies Division, Lawrence Berkeley National Laboratory
Berkeley, California 94720-8168, USA

[b]Division of Engineering, Brown University, Providence, Rhode Island 02912, USA

Cycling efficiency and rate capability of porous copper-coated, amorphous silicon thin-film negative electrodes are compared to equivalent silicon thin-film electrodes in lithium-ion batteries. The presence of a copper layer coated on the active material plays a beneficial role in increasing the cycling efficiency and the rate capability of silicon thin-film electrodes. Between 3C and C/8 discharge rates, the available cell energy decreased by 8% and 18% for 40 nm copper-coated silicon and equivalent silicon thin-film electrodes, respectively. Copper-coated silicon thin-film electrodes also show higher cycling efficiency, resulting in lower capacity fade, than equivalent silicon thin-film electrodes. We believe that copper appears to act as a glue that binds the electrode together and prevents the electronic isolation of silicon particles, thereby decreasing capacity loss. Rate capability decreases significantly at higher copper-coating thicknesses as the silicon active-material is not accessed, suggesting that the thickness and porosity of the copper coating need to be optimized for enhanced capacity retention and rate capability in this system.

[1] – International Society of Electrochemistry Active Member.
[2] – Present address: Department of Chemistry, University of California-Davis, One Shields Avenue, Davis, California 95616, USA.
[*] – Corresponding authors: Vijay A. Sethuraman (vj@cal.berkeley.edu); Venkat Srinivasan (vsrinivasan@lbl.gov); Phone: +1 510 764 4842/+1 510 495 2679

## 1. INTRODUCTION

The fully lithiated phase of silicon at room temperature is $Li_{15}Si_4$ [1], which translates to a maximum theoretical capacity of 3579 mAh/g for a silicon anode in a lithium-ion battery [2]. This is an order higher than graphite's theoretical lithium-intercalation capacity (*i.e.*, 372 mAh/g) [3,4]. This high capacity results in a significant increase in the energy density and specific energy of the cell (*ca.* 25 to 30%) [5]. The increased capacity combined with silicon's low discharge potential (< 0.5 V *vs.* Li/Li$^+$) makes it an attractive choice for use as negative electrodes in high energy lithium-ion batteries [6]. However, silicon and silicon-based negative electrodes exhibit huge volume expansion (*ca.* 270%) upon complete lithiation, and as a consequence, result in cracking and isolation of particles, leading in turn to cycling efficiency of less than 100% [[7,8]. This is because of the constant reforming of the solid-electrolyte-interphase (SEI) layer due to electrolyte reduction on freshly exposed surfaces [9]. An implication of this is the irreversible loss of lithium from the cell. Since cyclable lithium is a finite resource in a lithium-ion cell, this loss results in capacity fade. Nonetheless, silicon and silicon-based alloys are considered possible replacements to graphite-based negative electrodes used in today's commercial lithium-ion batteries [10]. Though the portable-electronics industry still remains the largest consumer of lithium-ion batteries, it is generally expected that these batteries will soon find a larger share in the automotive industry as hybrid electric vehicles (HEV) and plug-in hybrid electric (PHEV) vehicles become mainstream. Though there is a multitude of problems that needs to be addressed adequately before lithium-ion batteries can be used in automotive applications, improving the cycling efficiency is among the major ones. Coatings and electrolyte-additives play beneficial roles in mitigating some of the problems inherent in alloy anodes [10] and are in the early stages of exploration.

Recently, Mitsui mining [11] announced the commercialization of their new silicon-based composite negative electrodes (SILX®) for the PHEV market. This composite electrode is made of silicon particles (*ca.* 2 µm) coated with a porous layer of copper (*ca.* 500 nm) and is said to provide 30–50% higher power and *ca.* 100% increased capacity compared to today's carbon-based negative electrodes. The use of copper as an inactive but beneficial additive material in various forms (other than as substrate or current collector) in lithium-ion battery electrodes has been reported in the literature [12,13,14,15,16]. Kim *et al.* reported the improvement in cycleability of composite silicon negative electrodes after surface-modification by electroless-copper deposition [17]. They attributed the better cycling efficiency to conductivity enhancement from copper, and the better long-term cycleability to the formation of a $Cu_3Si$ alloy when the electrodeposited sample was annealed at high temperatures. Recently, Chiu *et al.* [18] reported the electrochemical performances of amorphous-silicon thin-film electrodes modified with copper nano-dots on their surface and showed lower capacity fade on these electrodes compared to equivalent silicon thin-film electrodes.

In this study, we evaluate the role of porous copper thin-film coating on the cycling characteristics of lithiated silicon thin-film electrodes. Cycling-efficiency and rate-capability data is not readily available in the literature for copper-coated-silicon negative electrodes (Si-Cu), and the objective of this study is to evaluate these metrics on copper-coated, silicon thin-film electrodes with thicknesses ranging from 10 nm to several hundred nanometers. In particular, the charge (lithiation) and discharge (delithiation) capacities, first cycle and steady-state cycling efficiency and rate capability of the copper-coated thin-film silicon electrodes are compared to equivalent silicon thin-film electrodes without copper coating. First, we evaluate if a copper underlayer plays a role in enhancing the cyclability of amorphous silicon thin-film electrodes. We then study the characteristics of the copper-coated silicon thin-film electrode during charge and discharge in a half-cell configuration (*i.e.,* against $Li/Li^+$ reference electrode).

## 2. EXPERIMENTAL

*2. 1. Electrode fabrication*

Thin copper discs (15.87 mm diameter, 0.3 mm thick) were used as substrates for electrode fabrication. Silicon thin films were prepared by RF-magnetron sputtering (Edwards Auto 306 DC and RF Sputter Coater) of a silicon target (3" diameter disc, 99.995% Si, Plasmaterials Inc., Livermore, CA) at 200 W power and at a pressure of 0.667 Pa of Argon (99.995%). Copper thin films were prepared by DC-sputtering of copper target (3" disc, 99.995%, Super Conductor Materials Inc., Suffern, New York) at 100 W and a pressure of 0.013 Pa of Argon. A 300 nm copper thin film (*i.e.,* a Cu underlayer) was first sputtered onto the copper disc followed by the deposition of 500 nm silicon film. The cell does not cycle very well without this copper underlayer (see Figure 3). For the copper-coated silicon electrodes, copper films of various thicknesses were sputtered on top of the 500 nm silicon film. Deposition rates for silicon and copper films were 125 and 75 nm/hr respectively.

*2. 2. Film thickness and composition measurements*

Film thicknesses were measured using an Alfa-Step IQ Surface Profiler. Calibration curves relating the sputter-deposition time to the film thickness were prepared (10-1000 nm range), and were used to arrive at proper deposition rates for silicon and copper. The thicknesses were also verified by cross-sectional scanning electron microscopy (SEM, JEOL JSM-6340F Field Emission Scanning Electron Microscope at 5 kV and 5mm working distance) images. Thickness values thus measured were used to calculate the mass of the film and hence its capacity (on a weight basis). The elemental composition of each of the layers was verified using Energy dispersive X-ray (EDX) spectroscopy (Genesis XM2 microanalysis system, EDAX Inc., Mahwah, New Jersey). The long-range order of the silicon surface was analyzed by Raman microscopy (Labram, ISA Groupe Horiba) with a helium-neon laser ($\lambda = 632.8$ nm) at 1 mW power as the excitation source. Finally, grain-size and short-range order of the silicon films were obtained from transmission electron microscopy data (Philips CM200/FEG at 200 kV).

*2. 3. Cell fabrication*

To ensure the complete removal of residual moisture, the sputtered electrodes were baked at 120 °C for 24 hours in vacuum (*ca.* 380 mm Hg). An electrode was then transferred into a glove-compartment without exposing it to air, and assembled into a 2032 coin cell configuration (*i.e.,* 20 mm diameter and 3.2 mm total thickness, National Research Council, Canada) under Ar atmosphere with a lithium-metal counter and reference electrode, and a woven Celgard 2500 separator (diameter = 1.9 cm, Celgard Inc., Charlotte, North Carolina). 1.2 M lithium hexafluorophosphate in a mixture of ethylene carbonate and diethyl carbonate (EC:DEC, 3:7 by % wt.) and 10% fluoro-ethylene carbonate (FEC) was used as the electrolyte (Novolyte Technologies, Independence, Ohio). The FEC additive increases the cycling efficiency of silicon anodes, probably due to the formation of a stable solid-electrolyte-interphase (SEI) layer [19,20]. A schematic with the coin-cell components is shown in Figure 1.

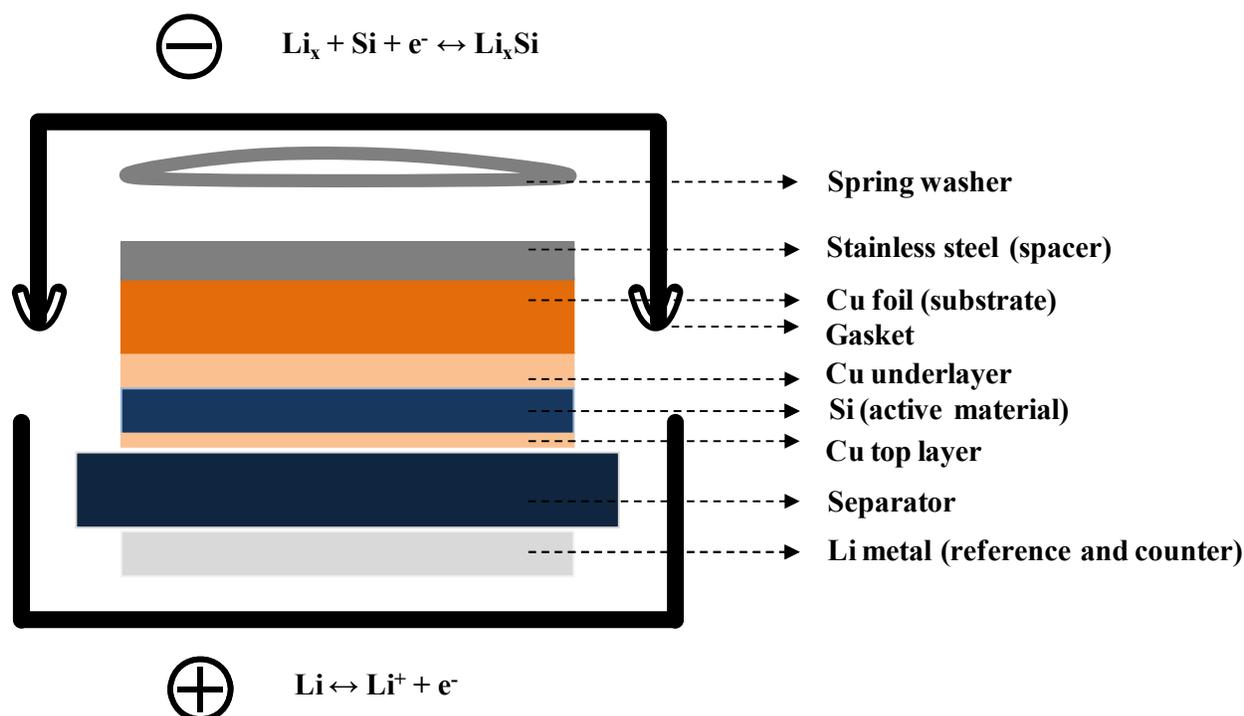

*Figure 1: Schematic of the coin-cell assembly used in this study with the constituent components and reactions at each electrode. Lithium metal was used as reference and counter electrodes. The positive and negative electrode reactions are shown. Note that this schematic is not drawn to scale.*

2. 4. Galvanostatic cycling

Electrochemical measurements were conducted in an environmental chamber at 23°C (±1°C) using a Solartron 1480A MultiStat system (Solartron Analytical, Oak Ridge, Tennessee), and data acquisition was performed using Corrware (Version 2.8d, Scribner Associates Inc., Southern Pines, North Carolina). The cell was cycled galvanostatically at 25 μA/cm$^2$ (geometric area, *ca.* C/8 rate; C/8 rate corresponds to a current allowing a full discharge in 8 h) total current between 0.01 and 1.2 V *vs.* Li. A lower cut-off potential of 10 mV *vs.* Li/Li$^+$ was chosen to prevent possible lithium deposition. Data acquisition rate was 1 Hz for all the electrochemical experiments. Open-circuit-potential relaxation for five minutes followed each charge and discharge steps. The input impedance of the instrument was 12 GΩ and the current due to the open-circuit potential measurement was negligible.

*2. 5. Rate capability studies*

Rate-capability experiments were conducted on cycled cells that had reached maximum, steady cycling-efficiencies, usually after 10-15 charge/discharge cycles. Rate capabilities of silicon, and copper-coated silicon thin-film electrodes were carried out such that charging was done at 25 µA/cm$^2$ (*ca.* C/8 rate), and discharging was performed at different rates from C/8 up to 3C. Both charging and discharging were conducted between 1.2 and 0.01 V *vs.* Li. Since the discharge capacities at higher rates were lower than charge capacity at C/8 rate, a one-hour open-circuit-potential relaxation was performed subsequent to discharging at higher rates, followed by a slower discharge at C/8 rate (until the system reached 1.2 V *vs.* Li/Li$^+$). This ensured that the electrode returned to a reasonably low state-of-charge (SOC) before the beginning of each rate experiment.

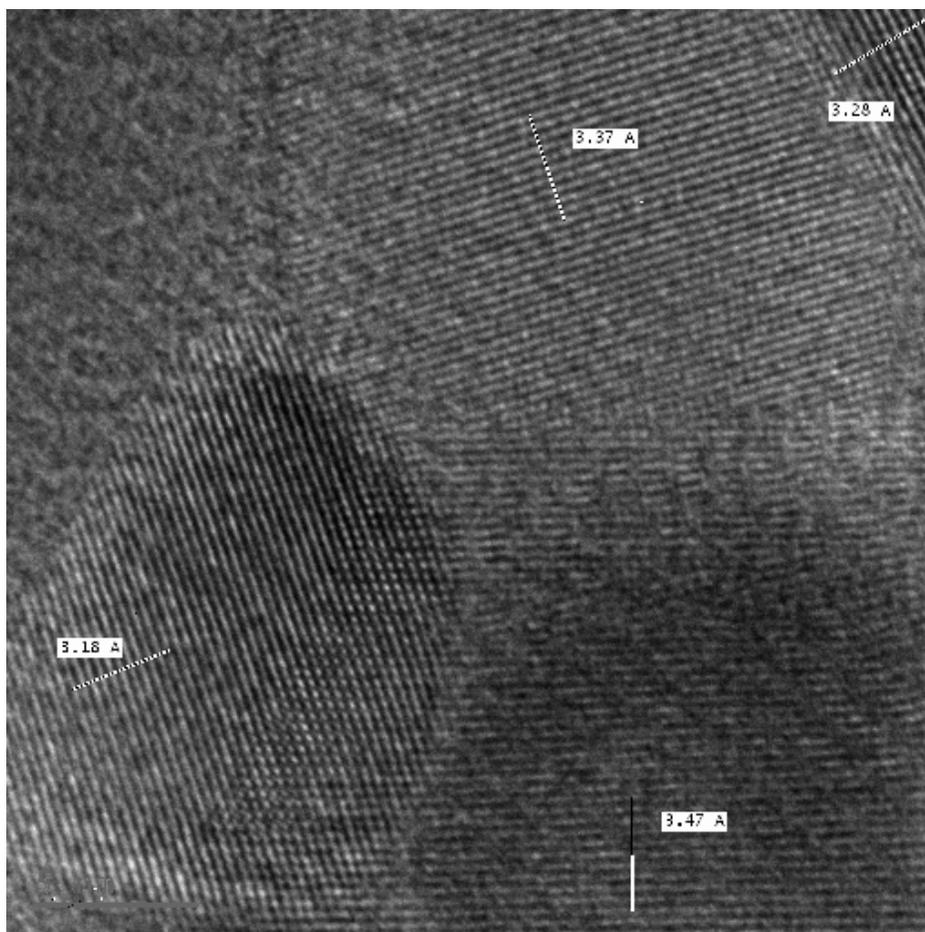

*Figure 2: High-resolution transmission electron micrograph of magnetron-sputtered silicon film. The length of the scale-bar is 5nm.*

## 3. RESULTS

*3.1. Thin-film characterization*

Transmission electron micrograph in Figure 2 shows that the sputtered silicon films were made of highly ordered 20-30 nm crystallites. SEM images (not shown) indicate that the copper films are continuous; it is possible for electrolyte-contact with the silicon film underneath the copper film *via* the edges. Besides, the porosity of these films was not characterized. The electron diffraction pattern (not shown) also confirms the nano-crystallinity of the silicon films. Raman spectroscopy data (not shown) suggests that these films are amorphous and nanocrystalline.

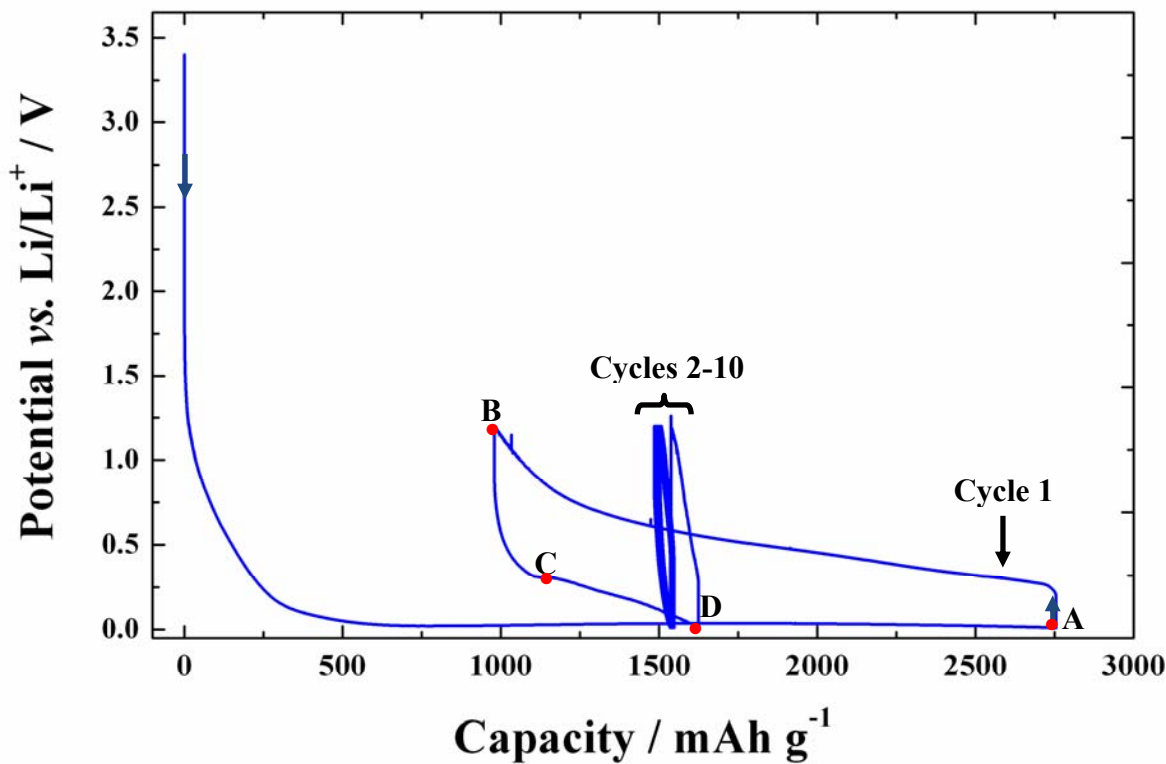

*Figure 3: Cell potential vs. capacity curves for lithiation and delithiation of a Si-Cu (40 nm) thin-film electrode without the 300 nm copper underlayer. The arrows show the cycling direction. The electrode was cycled at a C/8 rate between 1.2 and 0.01 V vs. Li/Li$^+$.*

Figure 3 shows the charge/discharge cycling data obtained from a Si-Cu (40 nm) electrode without the copper underlayer (*i.e.*, Si thin-film sputtered directly on the current collector). Without a copper underlayer, the electrode fails in the middle of the second cycle, and the electrode capacity falls to very low values in subsequent charge/discharge cycles. Such a dramatic failure may have occurred due to a significant loss of electronic-contact between the active material and the current collector. Beaulieu *et al.* [7] have shown that amorphous silicon thin-film electrodes undergo reversible and uniform shape and volume changes. They have also reported that surface-crack initiation in silicon thin-film electrodes occurs shortly after the beginning of first-cycle delithiation; such cracks propagate and widen when delithiation is continued till its end [9].

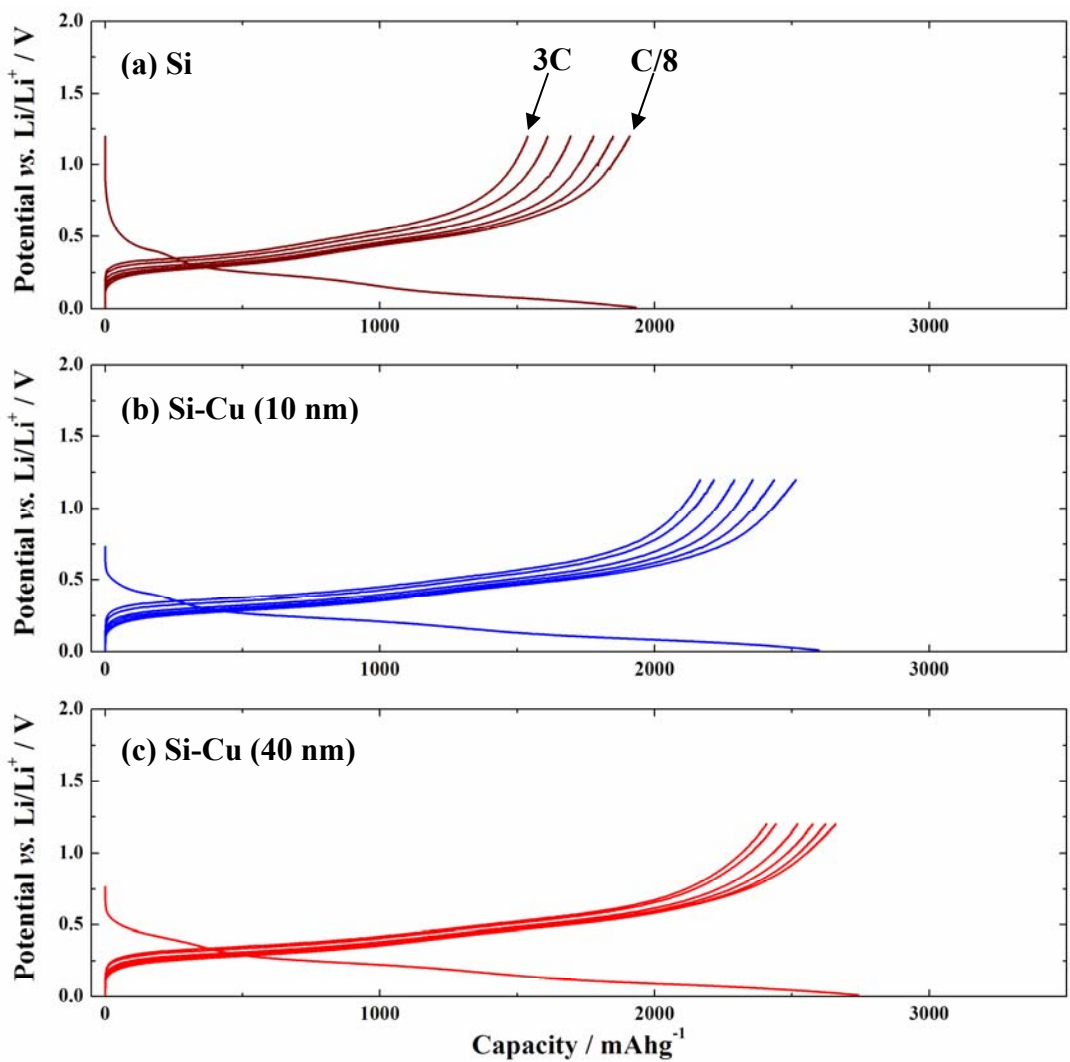

*Figure 4: Results from the rate-capability experiments on silicon and copper-coated silicon thin-film negative electrodes. Charging was done galvanostatically at 25 µA/cm$^2$ (ca. C/8 rate) with a lower cut-off potential of 0.01 V vs. Li/Li$^+$ and discharging was done at various rates from C/8 to 3C with an upper cut-off potential of 1.2 V vs. Li/Li$^+$. The intermediate rates are C/4, C/2, C and 2C.*

The silicon electrode also undergoes cycles of compressive and tensile stresses, respectively upon lithiation and delithiation [21]. In Figure 3, the electrode is under compressive stress (due to substrate constraint) until point A; the stresses become tensile during delithiation from point A to B; the stresses become compressive again during lithiation from point B to D. Until point B (*i.e.*, for one full charge/discharge cycle), the performance of this electrode is similar to that of an equivalent 40-nm coated silicon electrode with a copper underlayer (*i.e.*, the first cycle efficiencies and capacities are comparable). It is during the second cycle lithiation (*i.e.*, from B to C through to D), the performance of this electrode departs from that of an electrode with a copper underlayer. This is possibly due to delamination of the electrode from the substrate while it is under compression during lithiation.

We believe that delamination occurs during second-cycle lithiation and not during first-cycle lithiation due to the following reason: during the first cycle lithiation, the thin-film electrode is a monolithic structure and devoid of cracks (*i.e.*, interfacial cracks) and the energy-release rate required to delaminate a crack-free film is high. Based on Beaulieu *et al.*'s data [see figure 7 in ref. 7], it is reasonable to assume that the electrode starts to crack (*i.e.*, surface cracks which propagate rapidly to the Si/Cu interface whereupon they continue as interface cracks) during the first cycle delithiation, and at the end of delithiation (*i.e.*, at point B), the electrode is no longer a monolithic film but is made of islands of silicon with interfacial cracks. Upon subsequent lithiation, the substrate is no longer constraining the entire film but is constraining individual islands from expanding. The average energy release rate required for delaminating individual islands (or a severely cracked film) is considerably smaller than that of a monolithic, crack-free film. The presence of a sputtered copper film as an underlayer ensures proper adhesion between the film and the current collector, and prevents premature delamination. Furthermore, sputtered, amorphous copper films can sustain much larger elastic deformations than equivalent crystalline/bulk metallic structures [22,23]. As a result, the copper underlayer provides better resistance to delamination between the active material and the current collector at higher electrode stains, ensuring reversible cycling. The surface roughness of the current collector also plays a role in thin-film delamination. Since copper underlayer ensures better cyclability, subsequent studies on copper-coated silicon electrodes were all made on cells fabricated with this underlayer.

*3.2. Rate capability studies*

The delithiation curves corresponding to various rates from C/8 to 3C for Si, Si-Cu (10 nm), and Si-Cu (40 nm) are shown in Figure 4a-c, along with a lithiation curve at C/8 rate. Note that these data are obtained from well-cycled electrodes, typically after 10-15 charge/discharge cycles. The lithiation capacities at C/8 rate are different for these three electrodes because of capacity fade (*i.e.*, the initial lithiation capacities were similar for all three cells). The discharge capacity at any given rate is higher for a Si-Cu (40 nm) cell than the other two electrodes. We believe that the porous copper coating acts as glue that binds the electrode together, and prevents the electronic isolation of silicon particles. Being more ductile than silicon, copper possibly prevents the exposure of fresh surfaces during cycling, thereby minimizing loss of cyclable lithium due to solid electrolyte interphase (SEI) formation. While the former plays a role in increasing the conductivity of the silicon electrode and capacity retention, the latter plays a role in minimizing capacity fade and increasing the cell's cycling efficiency. Figure 4 also shows differences in polarization losses with increasing discharge rates. For example, the potential drop between C/8 and 3C rates at 50% state-of-charge is slightly lower for the Si-Cu (40 nm) electrode (in Figure 4c) compared to that of a Si electrode (in Figure 4a). This could possibly be an indication of the increased electronic conductivity of the film due to the copper coating.

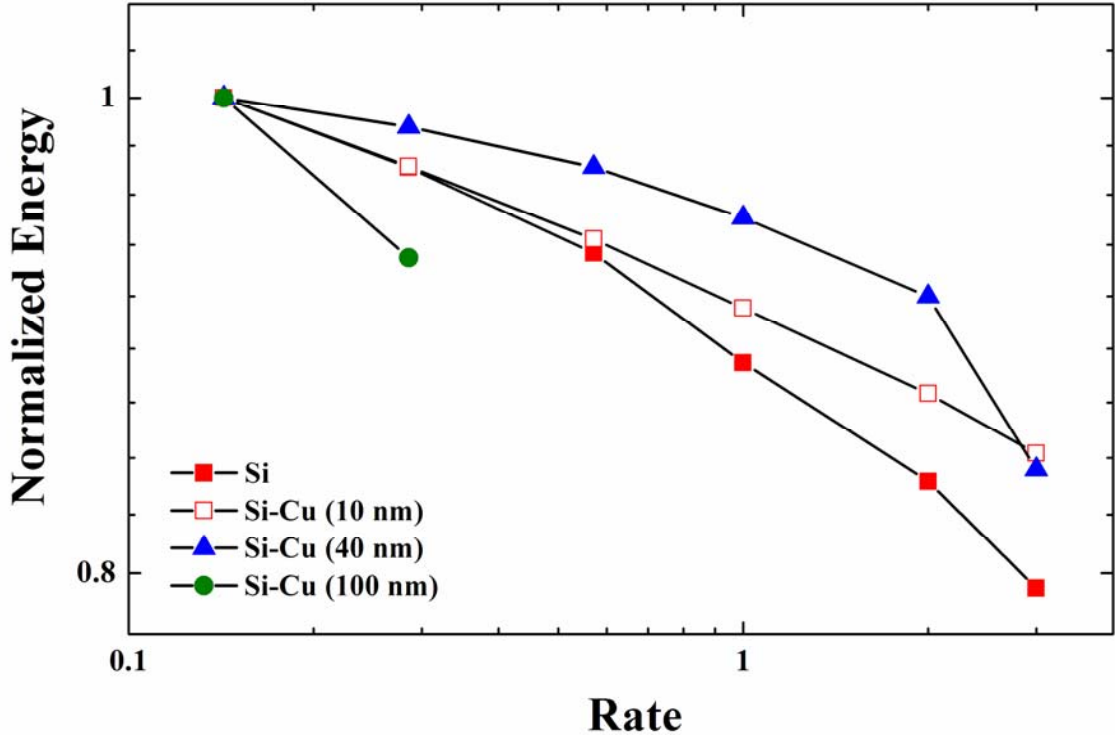

*Figure 5: Energy obtained at various discharge rates for silicon and copper-coated silicon negative electrodes are normalized based on energy obtained at ca. C/8 rate. A positive-electrode potential of 3.8 V vs. Li/Li$^+$ was used to arrive at the cell energy (Equation 1).*

Comparing the rate-capability of electrodes with disparate capacities was done by comparing the available cell energy normalized with respect to cell energy obtained at a low-discharge rate. The available cell energy was calculated using a constant cathode potential of 3.8 V *vs.* Li/Li$^+$ using the discharge curves obtained at various rates. This is given as,

$$Energy = \int_0^1 \{I\,[3.8 - V_a^d(z)]\}\,dz \qquad [1]$$

where $I$ is the discharge current, $z$ represents the SOC of the battery and varies from 0 (*i.e.,* at fully discharged state) to 1 (*i.e.,* at fully charged state), and $V_a^d(z)$ represents the anode discharge potential *vs.* Li/Li$^+$ as a function of $z$. Figure 5 shows the available energy for different discharge rates normalized against available energy obtained a C/8 rate. The Si-Cu (40 nm) electrode delivers higher energy at low to moderate discharge rates up to 2C but dips below Si-Cu (10 nm) electrode at 3C rate, possibly because of transport limitation. However, the rate-capability of copper-coated electrodes (except 100 nm) is better than that of an uncoated silicon electrode. The poor rate capability of a Si-Cu (100 nm) even at low rates is due to transport limitations, possibly due to blocking of the silicon from the electrolyte.

Silicon electrodes with higher copper thicknesses (*i.e.*, > 100 nm) were also studied; however their cycling and rate capability data is not shown here because of their poor performance. It has been reported that lithium transport through metallic-copper film occurs with a diffusion coefficient as high as $10^{-6}$ cm$^2$/s [24]. However, the capacities obtained for cells with copper thicknesses greater than 100 nm, when cycled galvanostatically at 25 μA/cm$^2$, were too low (*i.e.*, less than 200 mAh/g). Though the silicon electrodes with higher copper-coating thicknesses cycled well at extremely low rates, their performances were of no practical significance at rates of practical importance. Transport limitations appear to play a role in this result – *i.e.*, the copper film acts as a barrier for the lithium ions and as a consequence, not all silicon active material is effectively utilized. It is possible to tailor the porosity of such metallic coatings such that the weight of the inactive coating material is minimized (to maintain the energy density) while allowing for higher power.

*3.3. Cycling efficiency and capacity fade*

Cycle-to-cycle loss of lithium due to the constant reformation of SEI layer on fresh surfaces formed upon electrode cracking is one of many problems faced in alloy anodes that exhibit large volume expansion upon lithiation. Because of this, even on a well-cycled electrode, between two cutoff potentials, the charge capacity is always greater than the discharge capacity. Consequently, the cycling efficiency, η, the ratio of discharge to charge capacity, is always lesser than 1. Though this is true for all lithium-ion battery electrodes, full-cell cycling efficiency of state-of-the-art lithium-ion batteries with graphite anodes is upwards of 0.9999 [25], and that of the best silicon anodes is only 0.9995 [11,26]. Though the difference between these two efficiency numbers appears to be insignificant, the respective capacity retention at the end of 500 cycles is 95% and 77.8%. This is because the capacity retention of a lithium-ion battery is proportional to $C_d(\eta)^N$, where $C_d$ represents the initial discharge capacity, and N, the total number of charge/discharge cycles. Cycling efficiency, and discharge capacities of Si and Si-Cu (40 nm) as function of cycle number evaluated at a C/8 rate are shown in Figure 6. Average cycling efficiency (excluding the first cycle) of Si-Cu (40 nm) electrode is 99.35%, and that for the Si electrode is 98.89%. With the FEC additive, the cycling efficiency of Si-Cu (40 nm) electrode reaches upwards of 99% in fewer number of cycles than the Si cell. While these efficiency numbers do not compare favorably with the commercial lithium-ion batteries, the purpose of reporting these is to show that there is a finite improvement (*ca.* 0.5%) due to copper coating.

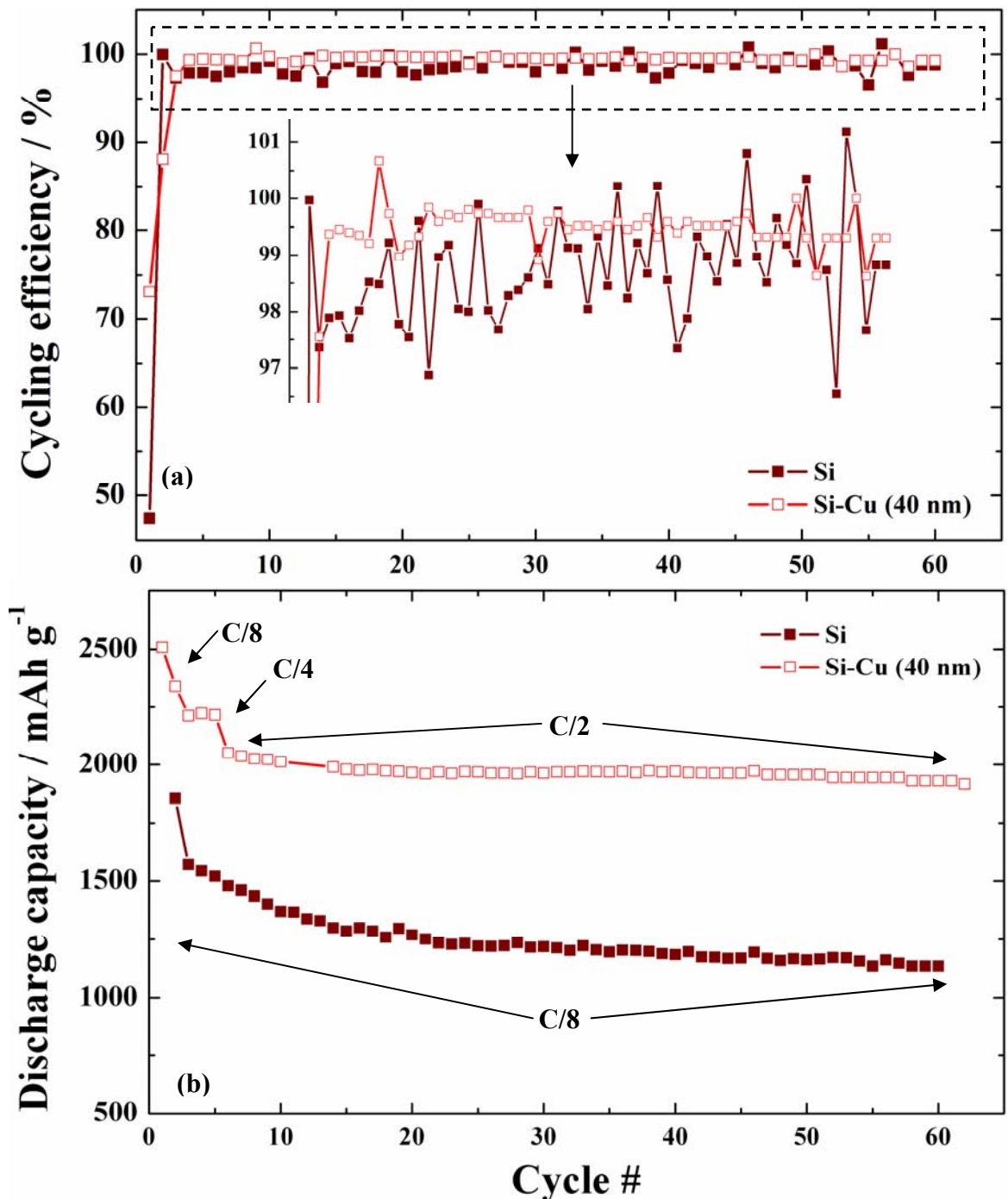

*Figure 6: (a) cycling efficiency and (b) discharge capacity is plotted as a function of cycle number for silicon and copper-coated silicon negative electrodes. These were evaluated in a coin-cell configuration with 1.2 M LiPF$_6$ in 3:7 EC:DEC and 10% FEC electrolyte and Li/Li$^+$ reference and counter electrodes. Charge/discharge cycles were done galvanostatically at 25 µA/cm$^2$ (ca. C/8 rate) for the Si electrode and between 25 and 100 µA/cm$^2$ (ca. C/8 to C/4 rate)*

*for the Si-Cu electrode. Respective cycling rates are indicated. Cycling was done between the potential limits of 0.01 and 1.2 V vs. Li/Li$^+$.*

Furthermore, the cycling efficiency holds steady over a number of cycles for the Si-Cu (40 nm) electrode whereas it fluctuates for the Si electrode (inset in Figure 6a), and decreases overall, possibly indicating that the Cu coating plays a beneficial role in holding the electrode together, and allowing it to expand and contract reversibly with lithiation and delithiation. The Si-Cu (40 nm) electrode thus exhibits lower capacity fade compared to the Si electrode (Figure 6b). The total capacity loss relative to the initial discharge capacity for the Si-Cu and Si electrodes at the end of the 60$^{th}$ cycle is 23% and 32%, respectively. The average discharge capacity of the Si-Cu (40 nm) electrode is 2002 mAh/g, much higher than that of the Si electrode's 1264 mAh/g. These performance metrics qualitatively indicate that Cu coating plays a beneficial role in improving the cycling efficiency of the lithiated silicon electrode.

*3.4. Implications and future studies*

These preliminary results have implications in furthering the scope of alloy-anode use in lithium-ion batteries for HEV and PHEV applications. While thin-film electrodes are excellent model systems, they have little practical importance in real-world batteries. Therefore, such studies have to be carried out in composite, porous electrodes that have immediate implications for commercial lithium-ion battery design and manufacturing. Silicon particles can either be coated prior to composite electrode fabrication, or such coatings can be electroplated directly onto a composite electrode. Nonetheless, the results presented in this article indicate that a metallic coating on silicon active material does play a beneficial role in increasing the capacity of the silicon electrode, retaining such capacity over a larger number of cycles, and providing increased rate capability. These positive attributes translate to a higher energy or a better range, longer life and higher power in PHEV and HEV batteries, respectively. Preliminary *in situ* stress measurements [21] on copper-coated silicon thin-film electrodes indicate that there are additional mechanical benefits compared to equivalent uncoated electrodes. The mechanics of metallic coatings on alloy anodes are currently ongoing in our laboratory. Furthermore, since such coatings are inactive and do not participate in lithium storage reactions, the thickness and porosity needs to be optimized to minimize the added weight while allowing for the benefits reported in this study.

**4. CONCLUSIONS**

Porous copper-coated silicon thin-film electrodes fabricated by magnetron-sputtering were evaluated for cycling efficiency, and rate-capability in lithium-ion coin cells, and compared to equivalent uncoated silicon thin-film electrodes. The presence of a copper layer between the electrolyte and the active material plays a beneficial role in increasing both the cycling efficiency and the rate capability of silicon electrodes in addition to minimizing overall capacity loss due to pulverization. We believe that copper, being more ductile than silicon, acts as a glue that binds the electrode together, and prevents the electronic isolation of silicon particles upon charge/discharge cycling, thereby decreasing capacity fade. The available energy and rate capability decreases significantly at higher copper thicknesses as the silicon active-material is not accessed, suggesting that the thickness and porosity of the copper coating need to be optimized for enhanced capacity retention and rate capability in this system, while minimizing the extra weight. The beneficial role of the copper-coating on the performance of composite silicon anodes, including the stress effects, is currently being investigated in our laboratory. The thickness and porosity of the copper coating are being optimized for capacity, cycling efficiency and rate capability in lithium-ion batteries.

## 5. ACKNOWLEDGEMENTS

The authors gratefully acknowledge the financial support from the Assistant Secretary for Energy Efficiency and Renewable Energy, Office of Vehicle Technologies, the United States Department of Energy, under contract no. DE-AC02-05CH11231 and the National Center for Electron Microscopy at LBNL, which is supported by the United States Department of Energy, under contract no. DE-AC02-05CH11231. The authors acknowledge the Micro-fabrication Laboratory at the Department of Electrical Engineering, University of California, Berkeley, for the fabrication of thin-film electrodes. The authors thank Xiangyun Song (LBNL) for assistance with electron-microscopy measurements. VAS gratefully acknowledges the support by the Materials Research, Science and Engineering Center (MRSEC) sponsored by the United States National Science Foundation, under contract no. DMR0520651. Helpful discussions with Professor Pradeep R. Guduru (Brown University) are gratefully acknowledged.